\documentclass[11pt]{article}

\usepackage[margin=1in]{geometry}
\usepackage{graphicx}
\usepackage{subcaption}
\usepackage{lineno}
\usepackage{xcolor}
\usepackage{glossaries}
\usepackage{gensymb}
\usepackage{appendix}
\usepackage{siunitx}
\usepackage{cleveref}
\usepackage{authblk}

\setacronymstyle{long-short}
\newacronym{pl}{PL}{photonic lantern}
\newacronym{mspl}{MSPL}{mode-selective photonic lantern}
\newacronym{itr}{ITR}{inverse taper ratio}
\newacronym{lp}{LP}{linearly polarized}
\newacronym{dcf}{DCF}{double-clad fiber}
\newacronym{na}{NA}{numerical aperture}
\newacronym{oam}{OAM}{orbital angular momentum}

\title{Modal response sensitivity to polarization across photonic lantern architectures}

\author[1,2]{Rodrigo Itzamn\'a Becerra-Deana}
\author[1,2]{Joseph Lamarre}
\author[1]{Rapha\"el Maltais-Tariant}
\author[1]{Adam Zolnai}
\author[1,2]{Nicolas Godbout}
\author[1]{St\'ephane Virally}
\author[1,2]{Caroline Boudoux\thanks{caroline.boudoux@polymtl.ca}}

\affil[1]{Polytechnique Montr\'eal, 2500 Chemin de Polytechnique, Montr\'eal, QC H3T 1J4, Canada}
\affil[2]{Castor Optics, 361 Boulevard Montpellier, Saint-Laurent, QC H4N 2G6, Canada}

\date{}

\begin{document}

\maketitle

\begin{abstract}
This paper examines the polarization-dependent output of various types of 3-mode photonic lanterns fabricated using double-clad fibers.

We explore the sensitivity of the modal response across several types of photonic lanterns, from the fully symmetric and strongly coupled structure of regular photonic lanterns to the fully asymmetric structure of mode-selective photonic lanterns.

We demonstrate the high sensitivity of the output of photonic lanterns with strong coupling between their ports to the polarization of the input state.

In contrast, ports with high isolation or low coupling, such as in mode-selective photonic lanterns, exhibit responses that are almost polarization independent.
\end{abstract}

\section{Introduction}

\Glspl{pl} are waveguide devices that enable adiabatic bidirectional light transfer between a set of single-mode waveguides and a single multimode waveguide~\cite{Leon-Saval:10}.
Some of these devices, \glspl{mspl}, deterministically map a given multimode domain to a tailored set of single-mode outputs~\cite{Li:21} and can be used as transverse mode multiplexers/de-multiplexers.
Both types of \glspl{pl} can exhibit broadband performance, up to 500~\unit{nm}~\cite{fontaine_photonic_2022,3PL_Becerra-Deana:25}, and small excess loss, less than 1~\unit{dB}~\cite{fontaine_photonic_2022,LEONSAVAL201746,birks_photonic_2015,Leon-Saval:10,CUI2023129550}.
Applications of \glspl{pl} range from their original role in astrophotonics, separating multimode starlight into different single-mode channels to process information in parallel~\cite{Birks:10}, to telecommunications, such as enabling high‑dimensional encoding by treating spatial modes as distinct information channels~\cite{LEONSAVAL201746}, and also biophotonics~\cite{deSivry-Houle:21,raphael_maltais-tariant_exact_2023,maltais-tariant_speckle_2023}.

Traditionally, the higher-order spatial modes associated with a \gls{pl} only act  as transient carriers, where the chief concern is achieving their combination and separation with minimal loss and crosstalk.

However, novel \gls{pl}-based techniques are emerging with the main objective of studying and using the optical properties of higher-order spatial modes\cite{velazquez-benitez_scaling_2018}.

Despite these advances, polarization is often an afterthought in the design and use of \glspl{pl}, or even when considering the wavelength response of the optical profile. The customary scalar \gls{lp}-mode framework treats each guided mode as having a single, degenerate transverse field component, thereby rendering the lantern nominally polarization-insensitive. Although polarization-maintaining \glspl{pl} have been demonstrated~\cite{10.1117/12.3057065},  the optical profile and final polarization orientation of the lobes remain undetermined by the fiber configuration, rotation, fusion, and tapering of several fibers. Consequently, the polarization of light injected into a \gls{pl} is rarely controlled, and any emerging polarization at the output is usually dismissed as an incoherent mixture. Ignoring polarization, however, can limit performance in several advanced applications. Even if fully polarization-maintaining \glspl{pl} remain impractical, deliberate manipulation of polarization offers a potent lever for controlling the multimode field and tailoring the polarization properties of higher-order modes. Adjusting the input polarization in the single-mode section can markedly alter, even rotate, the modal pattern delivered at the multimode output, as shown in~\cite{velazquez-benitez_optical_2018}.

In this paper, we experimentally study the influence of wavelength and, more importantly, polarization in the optical profiles of the three types of \glspl{pl}. We demonstrate how the input polarization determines the mode profile in non-selective devices, opening a new set of possible characteristics for applications. To do so, we use an optical system composed of an infrared camera, a polarization analyzer, polarization controllers, and different sources to allow us to observe and characterize the polarization response for a variety of \glspl{pl}. This paper also highlights how different \glspl{pl}, with respect to polarization, can have distinct applications. For example, \glspl{mspl} can be designed to be robust against polarization, ensuring that the optical profile remains unaffected while only the polarization of the lobes changes. In contrast, non-selective devices present the novel possibility of spatially encoding information using pure polarization through a single port. Each approach is important for different applications.

\section{Methodology}
\begin{figure}[h!]
        \centering
        \includegraphics[width=0.9\textwidth]{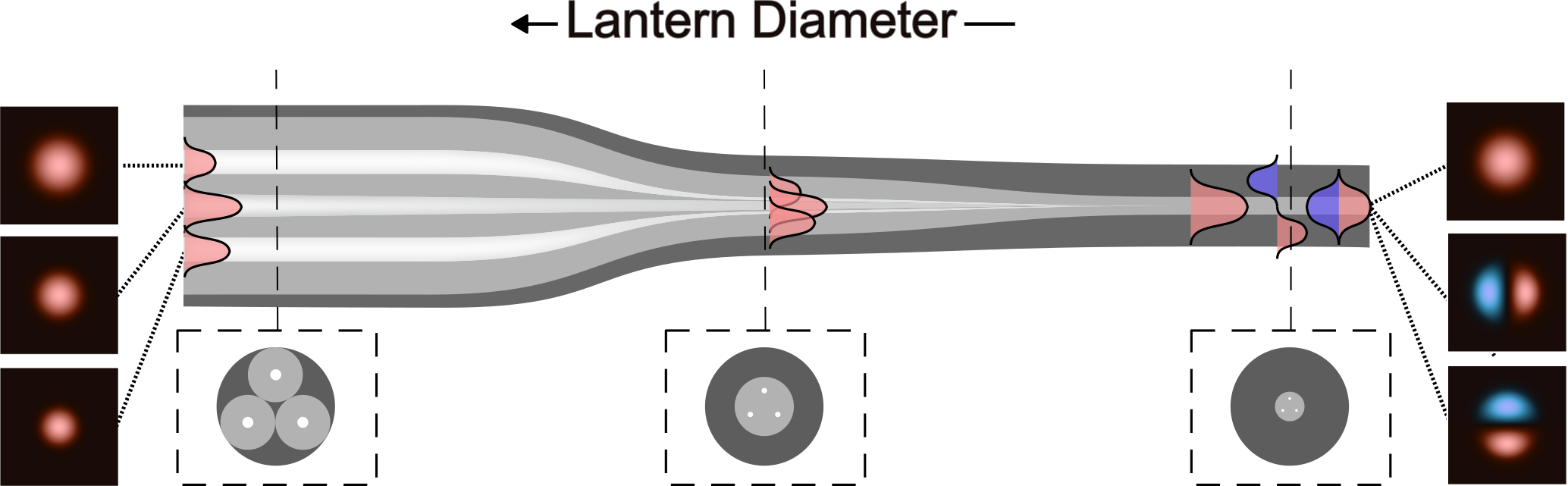}
        \caption{Modal evolution in a three-mode photonic lantern. Left: simulated near‑field intensity patterns of three input beams in the single-mode fibers. Middle, cross‑section of the fusion tapered process. Right: simulated near‑field intensity patterns profiles of the multimode output.}
        \label{fig:Lantern}
    \end{figure}

We analyzed the optical response of $3\times1$ \glspl{pl}, focusing on the coupling between modes under varying injection parameters, including wavelength and polarization state.  \Cref{fig:Lantern} shows a schematic of a three-mode \gls{pl} and the evolution of three single-mode inputs into the first three modes of the multimode region.

The \gls{pl} is a bundle of three single‑mode fibers, fused and adiabatically tapered into a single multimode section that supports the fundamental $\text{LP}_\text{01}$ and the doubly‑degenerate $\text{LP}_\text{11}$ set.  The figure does not purport to present only the case of an \gls{mspl}, where the correspondence between each input fiber and the three modes is one-to-one. In a regular \gls{pl} each single-mode input contributes to the amplitude to each mode.

The left-hand panel shows the near‑field amplitudes of the three single-mode fibers with exaggerated different mode field diameters. The right-hand panel shows the three isolated modes, $\text{LP}_\text{01}$ and even/odd $\text{LP}_\text{11}$. The middle section shows the adiabatic taper transition region. The vertical dashed lines highlight the geometric transition from the original bundle to the concentric multimode waveguide. 

The physical properties of the \glspl{pl} are, in part, determined by the original fiber bundle configuration. Representing different fiber types with different letters, the typical configurations include AAA or standard \glspl{pl} with three identical fibers, ABC or an \glspl{mspl}, and the hybrid configurations with either AAB or ABB~\cite{3PL_Becerra-Deana:25}.

For this project, all \glspl{pl} were fabricated using double-clad fibers with single-mode compatibility at 1550~\unit{nm} with each fiber, labeled A,B, and C, having a 9~\unit{\micro\meter} core diameter. The fibers are differentiated by their inner cladding diameters, namely 42, 32.3, and 19.6~\unit{\micro\meter} (fibers 2058I1, 2058K1, and 2058J1, Université Laval, Prof. Messaddeq, QC, Canada). The fabrication process is based on the heat and pull method presented in \cite{becerra-deana_mode-selective_2024}. The three double-clad fibers are inserted into a synthetic fused silica capillary tube (CV1012, Vitrocom, NJ, USA) prior to fusion to maintain the structural alignment of the fiber bundle.
    
\begin{figure}[htbp]
        \centering
        \includegraphics[width=0.8\textwidth]{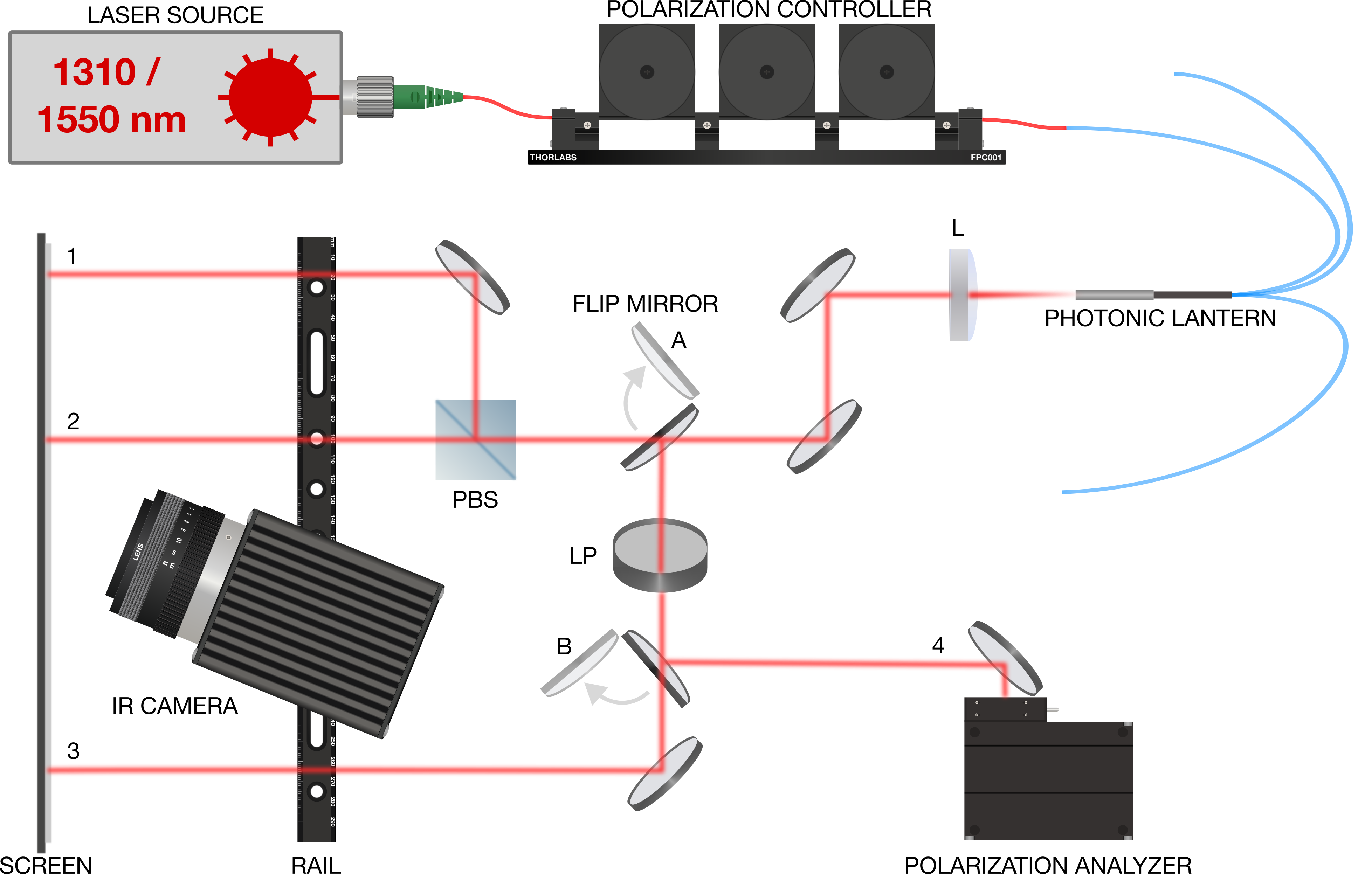}
        \caption{Schematic diagram of the optical system for polarization measurements, where PBS denotes a polarization beam splitter, P represents a polarizer, and L is a collimating lens. The two flip mirrors, A and B, lead to four possible paths toward screens 1, 2, and 3, as well as to the polarization analyzer in 4.}
        \label{fig:Schematics}
    \end{figure}

The modal response at the few-mode end of the \gls{pl} is analyzed under varying single-mode input polarizations and wavelengths, using the setup depicted in \cref{fig:Schematics}. This setup can accommodate two laser sources: a dual-wavelength narrow band source (HP 8153ASM, Hewlett-Packard, CA, USA) containing two single-mode lasers at 1300~\unit{nm} and 1550~\unit{nm}, and a broadband source (SL1310V1-10048, Thorlabs, NJ, USA) with a 100~\unit{nm} bandwidth centered around 1300~\unit{nm}. Both lasers are fiber-coupled to polarization controllers (FPC030, Thorlabs, NJ, USA ), which are then connected to a specific single-mode input of the \gls{pl}.

Different optical paths allow the observation of the optical profiles for various polarization states. The first two paths are the output of a polarization beam splitter, which separates the optical beam into its horizontal and vertical polarization components. Both outputs are then projected onto screens 1 and 2, where the modes are visualized using an infrared camera (SU320KTS-1.7RT, Goodrich, NJ, USA). This setup allows for the analysis of the polarization components of each optical mode and their evolution over varying input conditions. Flipping mirror A allows bypassing the polarizing beam splitter to a third optical path, projected onto screen 3 and visualized with the same camera, irrespective of the polarization state. Along this path, flipping mirror B allows for injecting the beam into path 4, which leads to a polarization analyzer (SK010PA-IR, Sch\"{a}fter + Kirchhoff, HH, DE) that measures the exact polarization state. A linear polarizer is inserted into this path to calibrate the polarization analyzer before acquiring measurements. The polarization analyzer indicates the position of the polarization state on the Poincaré sphere. The two main angles quantify, respectively, the ellipticity (linear polarization on the equator and circular polarization states at the poles) and the orientation of the ellipse's main axis. The radial component corresponds to the degree of polarization (DOP) as a percentage, from depolarized (0\%) to fully polarized (100\%). In the figures below, the various states are read on the polarization analyzer, and the numbers are converted into 2D pictorial representations (ellipses, circles, or lines).

Such a setup allows for straightforward control of the input parameters (polarization state and wavelength) injected into the \gls{pl}, as well as accurate measurement of the resulting few-mode output profile. Light is sequentially injected into all three ports across the three types of 3-mode \glspl{pl}, using five distinct polarization states and three wavelength configurations. The polarization states are: linear horizontal, linear vertical, left-- and right-handed circular, and one arbitrary elliptical state. The three wavelength configurations are: narrow-band at 1550~\unit{nm}, narrow-band at 1300~\unit{nm}, and broadband, i.e., 100~\unit{nm} around 1300~\unit{nm}. 

We coupled the light to the polarization analyzer in a free space configuration to ensure effective few-mode analysis of the polarization. All results are presented in the supplementary material, with some representative results in the main text.

\section{Results and Discussion}\label{sec:Results and Disucssion}

\begin{figure}[htbp]
        \centering
        \includegraphics[width=12cm]{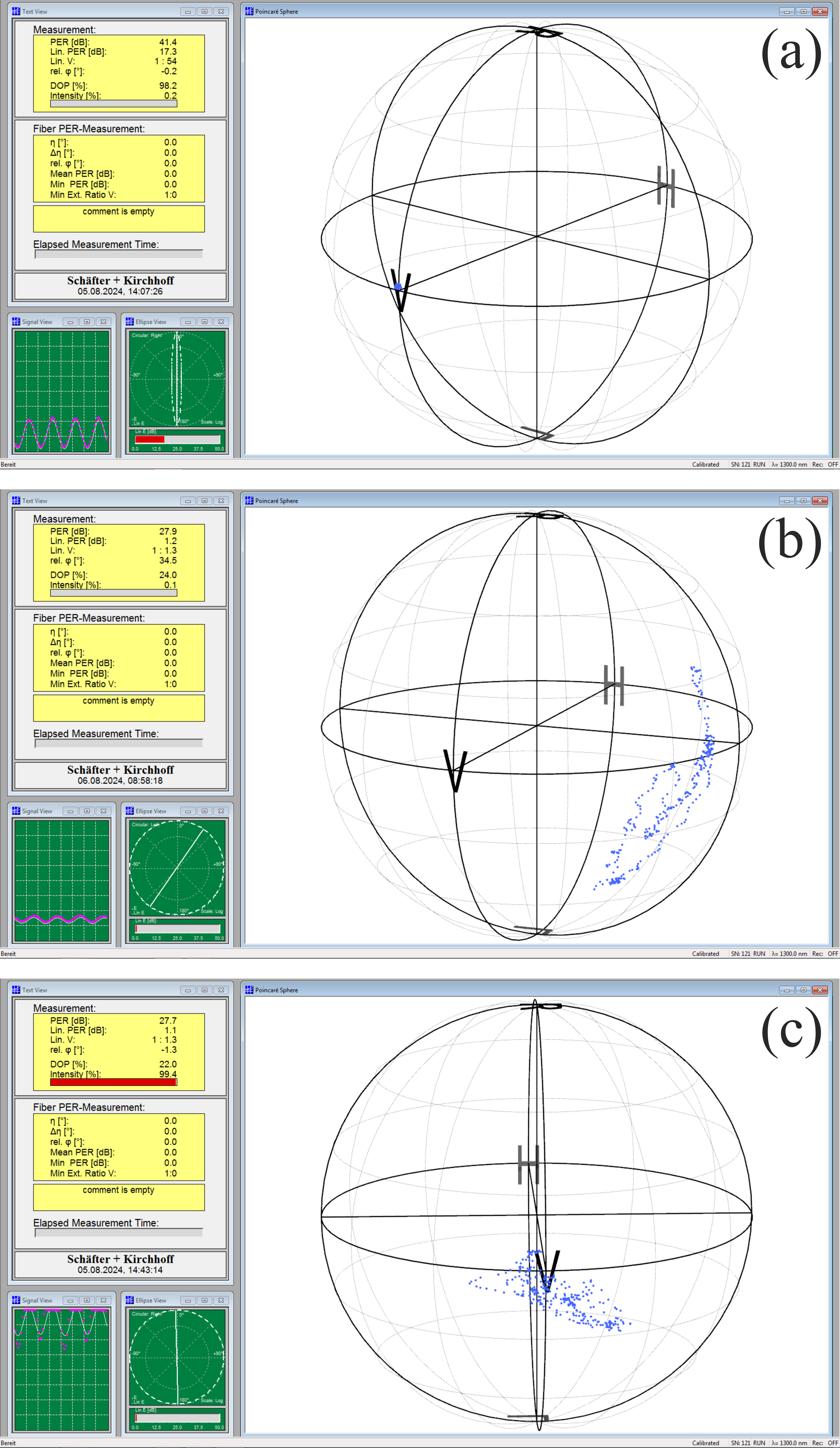}
        \caption{Measurements in the polarization analyzer. (a) The measurements done in the calibration, (b) with a AAA configuration at 1300~\unit{nm} short-band, and (c) AAA configuration at 1300~\unit{nm} broadband.}
        \label{fig:PolaError}
    \end{figure}

For the polarization analyses of each mode and profile, the four distinct paths illustrated in \cref{fig:Schematics} were used. The first three paths were used to obtain the optical profiles, while the fourth path specifically addressed the polarization response associated with each profile. It is important to note that the experiments were conducted using higher modes and the superposition of them rather than only the fundamental mode. To ensure accurate measurements, it was necessary to examine the output of the polarization analyzer and its behavior with these particular beams to identify any inconsistencies or inaccuracies in the polarization state measurements.
 Figure~\ref{fig:PolaError} presents three different cases of polarization measurements given in the Poincaré sphere by the polarization analyzer, and all the given values (PER [polarization extinction ratio], lin PER [polarization extinction ratio of linearly polarized light], $\varphi$ [azimuthal angle], rel $\varphi$ [relative azimuthal angle], DOP [degree of polarization], and Intensity -- see supplementary material) are observed in the first yellow square. Figure~\ref{fig:PolaError}(a) represents an example of the calibration measurement done to choose the output polarization, considering the polarizer before the polarization analyzer. The calibration was performed at the beginning of each port measurement, following the procedure described below. The port to be measured was illuminated so that light continued along path number 4. A polarizer was placed between the two flip mirrors (A and B in \cref{fig:Schematics}) to ensure that only vertical polarization reached the polarization analyzer. The power was maximized by adjusting the polarization controllers to achieve vertical polarization. In some cases, this was further verified by monitoring paths 1 and 2 to confirm that the polarization was predominantly vertical. As mentioned, the calibration was done with the supermode structures.

Additionally, Figure~\ref{fig:PolaError}(a) shows an adequate case measurement at 1300~\unit{nm} where each measurement in time, represented by a blue dot in the Poincaré sphere, falls into the same targeted polarization (vertical polarization in this case). Although the intensity in this case was low, the polarization was adequately captured, showing a DOP of 98.2.

 Figure~\ref{fig:PolaError}(b) displays measurements taken at 1300~\unit{nm} using an AAA configuration, and Figure~\ref{fig:PolaError} presents measurements taken at 1300~\unit{nm} using a broadband source with the same \gls{pl}. Figure~\ref{fig:PolaError}(b) and (c) illustrate cases where the polarization response varies significantly in time, making it difficult to determine. Even though calibration was performed for each experiment, in some cases, such as Figure~\ref{fig:PolaError}(b) and (c), at the moment of the measurements, the measurements were abrupt, as in these images.

Limitations of the current system demonstrate discrepancies in some aspects of the polarization measurements, as shown in Figure~\ref{fig:PolaError}. A possible explanation for the abrupt polarization measurements is the response of \glspl{pl} with strong coupling between ports (the repetition of at least one type of fiber AAA, AAB, ABB) where the injection port experiences significant power transfer and interactions with the other ports due to mode superposition, which can be affected by any change in phase. These structures suggest that the superposition of modes could generate partially unpolarized light. This behavior was also observed at 1300~\unit{nm}, as shown in the supplementary material. 

Furthermore, the polarization analyzer has a short-band response (a single wavelength was set for each measurement); however, broadband measurements were still taken and showed satisfactory results, which are detailed in the supplementary material.

\subsection{AAA configuration}
    
In all fiber configurations, the polarization state was controlled using the polarization controllers and monitored with the polarization analyzer. In this way, the measurements obtained from the polarization analyzer confirmed the polarization states observed in paths 1 and 2. Once the mode's polarization was fixed, the measurements were taken. Each of the following figures provides all the data from the four paths at different polarizations. However, when the polarization states are inconsistent, the figures display only the optical profiles corresponding to the three reliable paths. Figure~\ref{fig:AAA1550nm} illustrates the five random polarization states for each port at a wavelength of 1550~\unit{nm}. Each row corresponds to a type of polarization, while each pair of columns relates to its corresponding injection port, showing three optical profiles. The first column displays the composition power distribution of the polarization, and the second column shows the two orthogonal profiles, horizontal and vertical, respectively represented with the arrows at the top of the first row.

   \begin{figure}[htbp]
        \centering
        \includegraphics[width=0.55\textwidth]{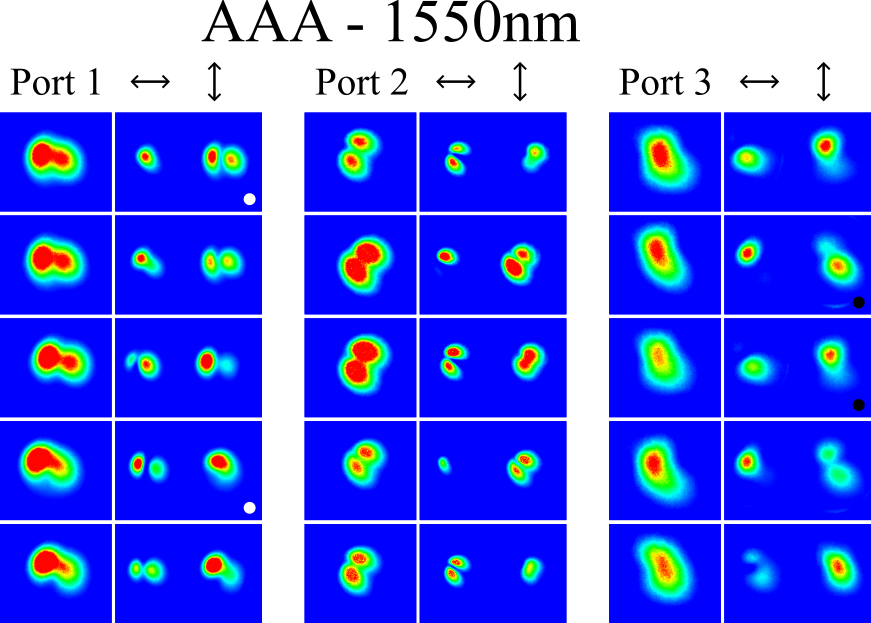}
        \caption{Optical intensity profile of each port of a standard 3-mode \gls{pl} at five tested polarizations. Each row represents a polarization, while each pair of columns corresponds to its associated port, displaying three optical profiles. The first column illustrates the composition of the polarization, and the second column shows the two orthogonal profiles: horizontal and vertical.}
        \label{fig:AAA1550nm}
    \end{figure}

The analysis of the optical profiles in Figure~\ref{fig:AAA1550nm} reveals a power fluctuation between the lobes, with more drastic responses observed in the orthogonal polarization modes view (columns with the horizontal and vertical arrows). Focusing on the first port, particularly in the first and fourth rows (both highlighted with a white dot), it appears that the $\text{LP}_\text{01}$ and $\text{LP}_\text{11}$ modes at those polarizations exhibit a form of degeneration that can be distinguished. This phenomenon is also evident at port 2 across all ports, where, in some cases, the generation of $\text{LP}_\text{11}$ is separated into horizontal or vertical polarizations, with the remaining power in the opposite polarization. Additionally, power fluctuations in the second and third rows of port 3 of the images of horizontal and vertical responses show significant changes spatially (both highlighted with a black dot). Additional examples of this phenomenon can be found in the supplementary material (Fig. S1, Fig. S2, and Fig. S3).

\subsection{ABB configuration}

\begin{figure}[htbp]
  \centering
    \begin{subfigure}[htbp]{\textwidth}
        \centering
        \includegraphics[width=0.7\textwidth]{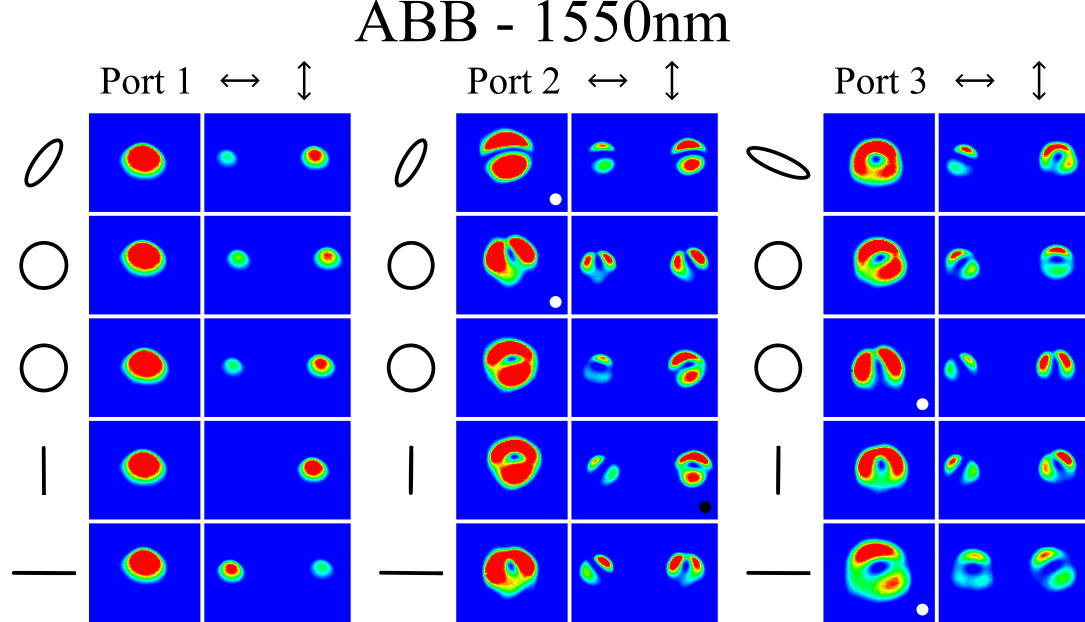}
        \caption{Hybrid \gls{pl} at 1550~\unit{nm}}\label{fig:ABB1550}
    \end{subfigure}%
\vspace{1em} 
    \newline 
    \begin{subfigure}[b]{\textwidth}
        \centering
        \includegraphics[width=0.7\textwidth]{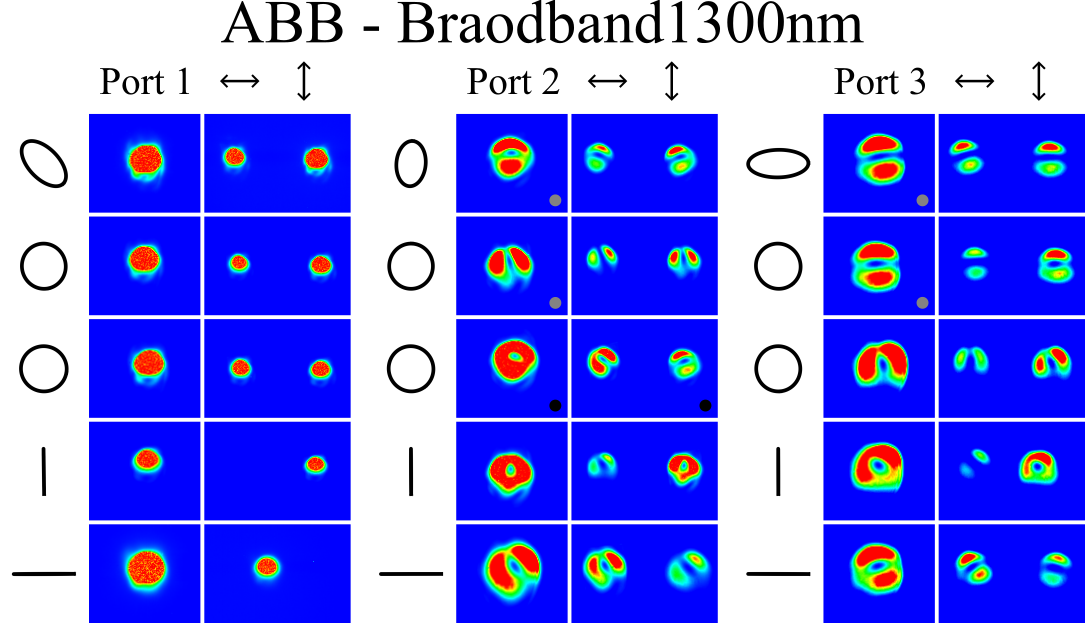}
        \caption{Hybrid \gls{pl} with a broadband source at 1300~\unit{nm}}\label{fig:PABBB}
    \end{subfigure}
    \caption{The optical intensity profile of each port of the same ABB \gls{pl} at two wavelengths (a) and (b). Each row represents a specific type of the five polarizations, while each pair of columns corresponds to its associated port, displaying three optical profiles. The first column illustrates the composition of the polarization, and the second column shows the two orthogonal profiles: horizontal and vertical.}
\end{figure}

The AAB and ABB configurations have comparable responses, strong coupling with some ports (the one that repeats the letter configuration), and high isolation in the ports that have a different fiber than the others. Their main difference lies in the superposition of modes generated by the few-mode section, which is related to the fiber characteristics and configuration\cite{3PL_Becerra-Deana:25}. Their response is a blend of the AAA configuration observed before and the ABC configuration. As a result, this section will solely focus on the ABB configuration for clarity. To maintain consistency in the analysis, the methodology for data acquisition and presentation mirrors that used in the AAA configuration. Figure~\ref{fig:ABB1550} presents the ABB configuration's polarization profile at 1550~\unit{nm} across its three ports. On the left side of each mode image is a polarization projection of the corresponding mode, obtained using the polarization analyzer's polarization values. Adjacent to that, two optical profiles represent the horizontal and vertical polarization states. On the other hand, Figure~\ref{fig:PABBB} shows the optical profile for the ABB configuration using a broadband source centered at 1300~\unit{nm} with the same layout as previously described.

ABB configuration (Figure~\ref{fig:ABB1550}) can support a donut shape through the superposition of the two modes using just a single fiber, as seen in the first row of the third port of Figure~\ref{fig:ABB1550}. In this specific instance, this type of performance can be observed at port 2 in the first and second rows of the images, divided by their polarization (both highlighted with a white dot), or even in port 3 in the third and fifth rows. Clearly, the transition from $\text{LP}_\text{11a}$ to $\text{LP}_\text{11b}$ is not yet complete, but this does not imply that a full transfer could have occurred in a different polarization state. Still, this mode-orthogonality transition suggests a potential method for spatially encoding information via pure polarization, which could occur in devices with strong mode coupling. Furthermore, examining the orthogonal modes in the fourth row of port 2 reveals the generation of two $\text{LP}_\text{11}$ modes with orthogonal polarizations (highlighted with a black dot), yielding results similar to some cases displayed in Figure~\ref{fig:AAA1550nm} of mode separation throughout polarization.

Additionally, Figure~\ref{fig:PABBB} shows the response of these devices using a broadband source, where instead of generating a superposition of all the modes into a donut shape or a Gaussian profile due to the summation of all optical profiles per wavelength, the response is similar to that in Figure~\ref{fig:ABB1550}, a separation of the modes. It is possible to force all wavelengths to generate a specific mode, enabling the creation of quasi-$\text{LP}_\text{11}$ modes using one port of a semi-mode-selective \glspl{pl}, as shown in the first two rows for ports 2 and 3, highlighted with a gray dot. Still, similar to Figure~\ref{fig:ABB1550}, there are cases where the doughnut shape could be generated with only one port, and more interesting to generate the two quasi-$\text{LP}_\text{11}$ modes with orthogonal polarization for several wavelengths, which is shown in the images with the black dot of Figure~\ref{fig:PABBB}.

In both scenarios, whether using a single-wavelength source or a broadband source, the response at port 1 is consistent in its profile. This is because port 1 breaks the structure's symmetry and creates a mode with high isolation relative to the others.

\subsection{ABC configuration}

\begin{figure}[htbp]
        \centering
        \includegraphics[width=0.7\textwidth]{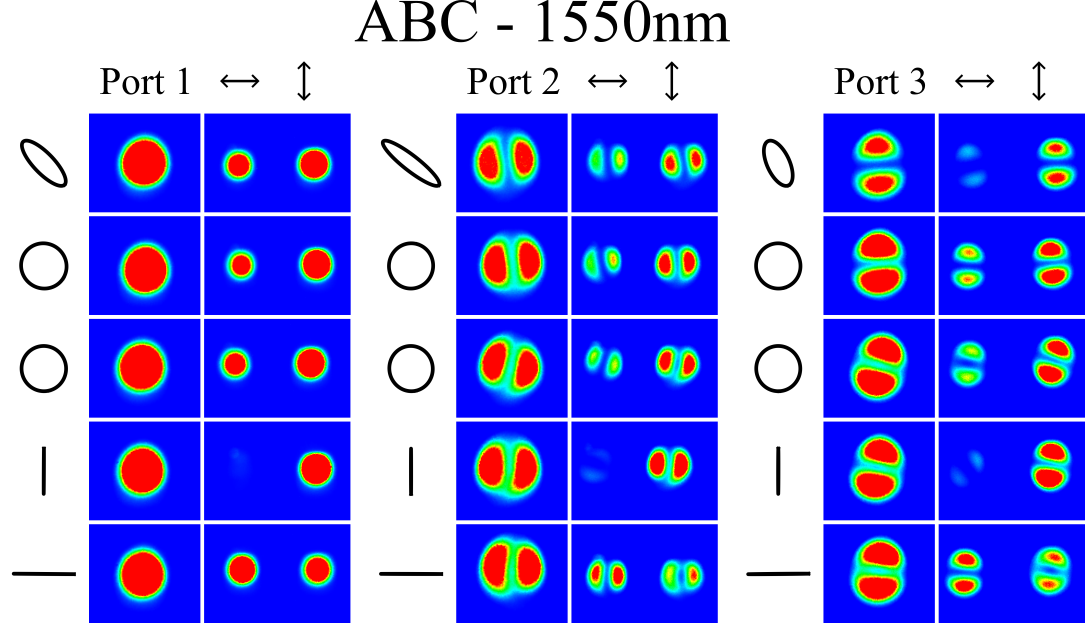}
        \caption{The optical intensity distribution for each port of two types of 3-mode \gls{pl} (a) and (b). Each row corresponds to a particular type of polarization, while each set of columns relates to its respective port, showing three primary optical power distributions. The first column represents the polarization composition, while the second column displays the two orthogonal profiles: horizontal and vertical.}
        \label{fig:ABC1500nm}
    \end{figure}

The last configuration considers the behaviour observed on port 1 of the ABB configuration, as each fiber is different and breaks the symmetry in all directions, generating an optical profile stable and quasi-invariant with respect to polarization. Figure~\ref{fig:ABC1500nm} shows the polarization profile at 1550~\unit{nm} for the ABC arrangement across its three ports. A polarization projection of the respective mode is on the left side of every mode image. Next to that, two optical profiles illustrate the horizontal and vertical polarization states.

    In contrast with all configurations, the ABC configuration exhibits quasi-spatial invariance to any wavelength and polarization injection, as shown in \cite{becerra-deana_mode-selective_2024}. Still, there was a case of discrepancy in the polarization that occurred when the polarization did not fully match the response in the optical profile after the beam splitter, as seen in Figure~\ref{fig:ABC1500nm} for horizontal polarization in the first port. In some cases, these differences could be due to some perturbation caused by the flip mirrors.

\begin{figure}[htbp]
        \centering
        \includegraphics[width=4.5cm]{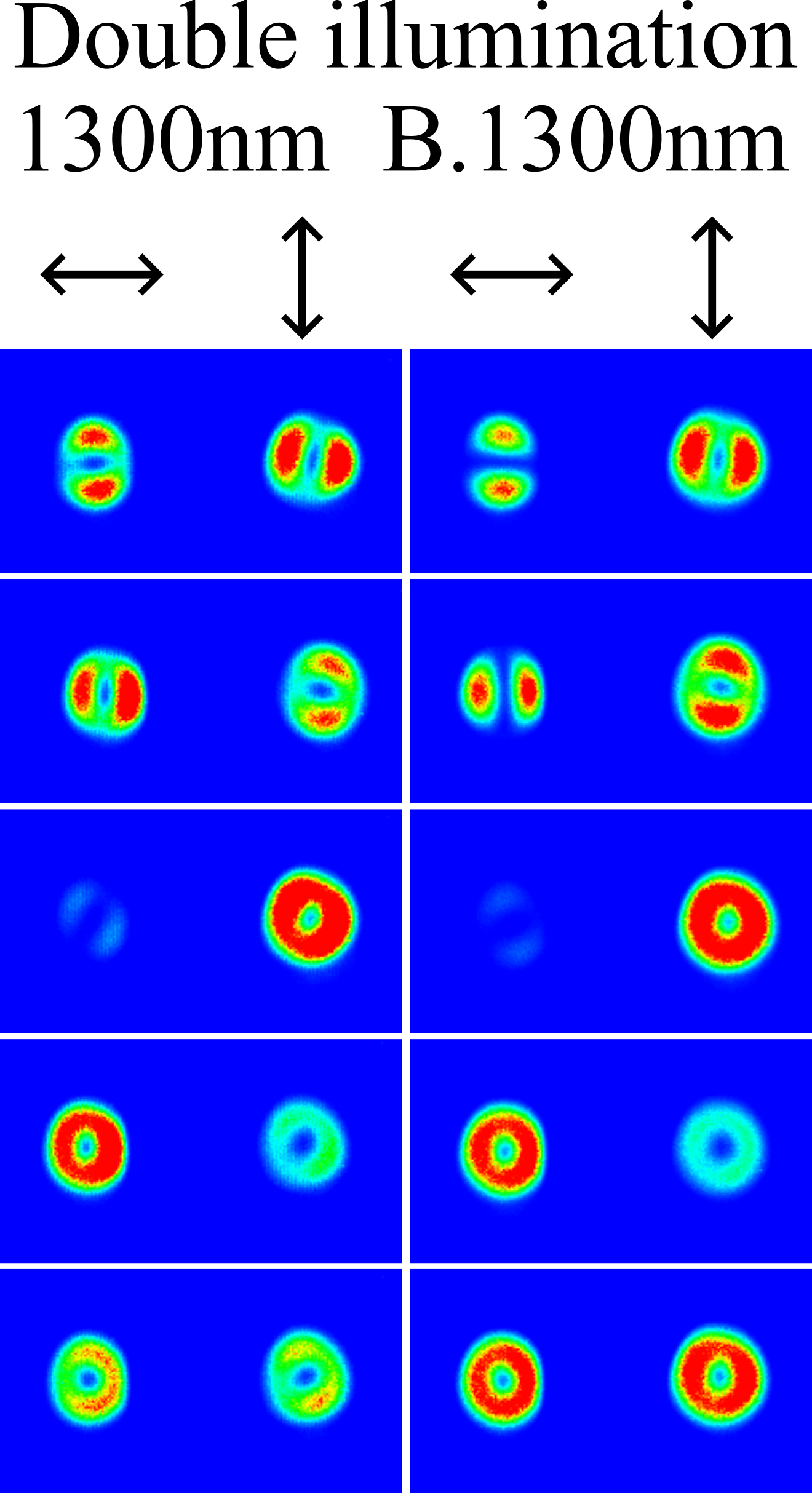}
        \caption{Optical intensity profile in a \gls{mspl} when the two $\text{LP}_\text{11}$ modes are illuminated at different polarization states. The left column shows the decomposition of the polarization into horizontal and vertical using a short bandwidth, while the right column employs a large bandwidth with the same configuration. Each row illustrates different polarization responses on the donut shape.}
        \label{fig:TwoLP11}
    \end{figure}

The ABC configuration can be used to achieve a superposition of both LP$_{11}$ modes with a 50/50 coupler incorporated after the laser and before the polarization controllers in the system of Figure~\ref{fig:Schematics}. Figure~\ref{fig:TwoLP11} illustrates the intensity profile of the two orthogonal polarization states when the two $\text{LP}_\text{11}$ ports of an ABC configuration are illuminated with two different light sources, a short-band at 1300~\unit{nm} (the left column) and broadband at 1300~\unit{nm} (the right column). This particular configuration had a significant optical path difference, allowing for the generation of a donut shape rather than an $\text{LP}_\text{11x}$ or $\text{LP}_\text{11+}$ mode~\cite{STED_Singlefiber_modesSum, Alarcon:23, CombEr,CUI2023129550}.

\section{Conclusion}\label{sec:Conclusion}

We conducted an extended characterization of the polarization of three types of \glspl{pl}. Understanding how each type of \gls{pl} responds to variations in polarization states and to broadband sources enables a more refined approach to designing and utilizing these devices for specific applications. The methodologies employed in this study successfully highlighted the differences in response depending on lantern configurations and polarization states, as evidenced by the integration of various laser sources. It was shown that configurations with strong coupling between each port (for instance, AAA, AAB, and ABB) exhibit significant variation in intensity profiles based on wavelength and polarization variations. 
We further showed that certain polarization states can generate quasi-$\text{LP}$ modes for both short-band and broadband responses. Each \gls{pl} may have distinct applications based on its specific performance. For instance, the AAA configuration is suitable for spatial multiplexing with a single port by changing polarization or wavelength. The ABC configuration, which is independent of wavelength and polarization, shows significant promise for specialized applications, such as biomedical imaging, where mode multiplexing and demultiplexing can enhance imaging~\cite{raphael_maltais-tariant_exact_2023}. 
\section*{Funding}
Caroline Boudoux and St{\'e}phane Virally acknowledge funding from the Mid-Infrared Quantum Technology for Sensing (MIRAQLS) project, supported by the European Union’s Horizon Europe research and innovation programme under grant agreement 101070700. Natural Sciences and Engineering Research Council (NSERC) of Canada grants \#RGPIN-2018-06151 (CB).

\section*{Acknowledgment}
The authors thank Mika\"el Leduc for his invaluable work on the fabrication and characterization setup.

\section*{Disclosures}
RIBD: Castor Optics Inc. (E). CB: Castor Optics Inc. (I). NG: Castor Optics Inc. (I). The other authors declare no conflicts of interest.

\section*{Data Availability}
Data underlying the results presented in this paper is not publicly available at this time but may be obtained from the authors upon reasonable request.

\section*{Supplementary Material}
\section{Polarization measurements}

This section includes measurements for four \glspl{pl}, five polarizations, and three light sources, detailing the optical profiles of composition, as well as horizontal and vertical polarization with their respective measurements from the polarization analyzer. The polarization analyzer provides several key measurements: the polarization extinction ratio (PER), which compares the power of two linear polarization states on a logarithmic scale. Another measurement is linear PER, representing the proportion of linearly polarized light to the total light detected, also expressed logarithmically. Additionally, $\varphi$ denotes the azimuthal angle indicating the alignment of the primary polarization axis. Lastly, the degree of polarization (DOP) indicates the fraction of light that is polarized. 

The measurements of the \gls{pl} are systematically organized by wavelength, providing a comprehensive overview of the optical profiles along with a detailed table of measurements for each port. Each figure is configured into three columns per port: 
\begin{itemize}

\item Column 1: Displays the 2D projection of the polarization, illustrating the orientation and distribution of the polarization states.
\item Column 2: Presents the optical profile, showcasing the intensity distribution across the beam for the respective port.
\item Column 3: Offers a dual representation of the optical profile, displaying the configurations in both horizontal and vertical orientations.
\end{itemize}

It is important to note that in certain instances, there may be a lack of polarization projection. This is attributed to abrupt measurement as discussed in the main paper. Each row corresponds to one of the five defined polarizations, ensuring clarity in the representation of polarization states across all ports.

Adjacent to the figures, a comprehensive table provides the polarization measurements, following the same sequence as the corresponding optical profiles for each port. This arrangement allows for easy cross-reference between the visual data and the numerical measurements, facilitating a thorough understanding of the \gls{pl}'s performance across different wavelengths and polarization states.

\subsection{AAA configuration}

\begin{figure}[h!]
        \begin{minipage}[b]{0.55\linewidth} % Adjust width as needed

        \includegraphics[width=1\linewidth]{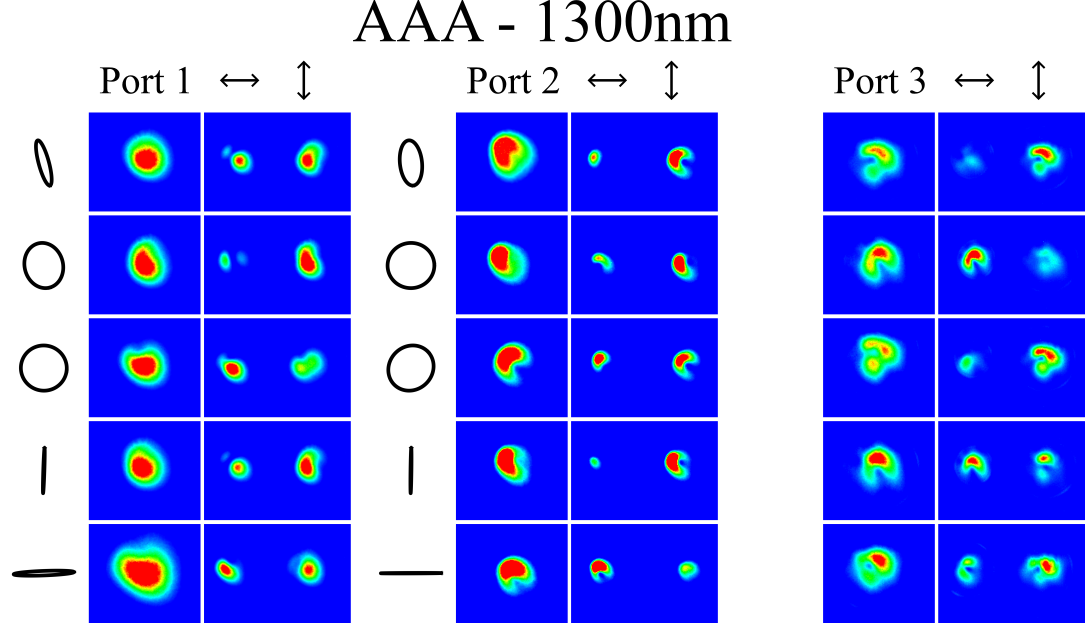}
        \caption{Optical profile response at five different polarizations of a standard \gls{pl} at 1300~\unit{nm}.}\label{fig:AAA1300nm}

        \end{minipage}
        \hfill % Adds horizontal space between minipages
        \begin{minipage}[b]{0.45\linewidth} % Adjust width as needed
\begin{center}
\setlength{\tabcolsep}{1pt}
\tiny
\begin{tabular}{ |c c|c c|c c| } 
 \hline
 \multicolumn{2}{|c|}{Port 1} & \multicolumn{2}{|c|}{Port 2} & \multicolumn{2}{|c|}{Port 3} \\

\hline\hline

PER [dB] & lin PER [dB] & PER [dB] & lin PER [dB] &PER [dB] & lin PER [dB] \\ 
$\psi$ [\degree] & DOP [\%] & $\psi$ [\degree] & DOP [\%] & $\psi$ [\degree] & DOP [\%] \\
\hline

13.11 & 8.23 & 6.18 & 5.10 & N.A. & N.A.\\ 
-74.73 & 89.06 & -83.97 & 90.88 & N.A. & N.A. \\
\hline

1.44 & 1.33 & 0.36 & 0.26 & N.A. & N.A. \\ 
-78.46 & 93.56 & 2.38 & 72.12 & N.A. & N.A. \\
\hline

0.20 & 0,20 & 0.74 & 0.68 & N.A. & N.A. \\ 
-7.72 & 96.54 & 42.00 & 91.71 & N.A. & N.A. \\
\hline

 32.04 & 10.56 & 33.70 & 8.96 & N.A. & N.A.\\ 
88.11 & 91.26 & 89.23 & 87.33 & N.A. & N.A.\\
\hline

22.54 & 13.20 & 41.27 & 6.52 & N.A. & N.A. \\ 
2.15 & 95.75 & 0.20 & 77.74 & N.A. & N.A. \\
\hline

\end{tabular}

        \captionof{table}{Measurements of a \gls{pl} for each port across five polarizations at 1300~\unit{nm}.}
        \label{tab:T1}
\end{center}
        \end{minipage}
    \end{figure}

\begin{figure}[h!]
        \begin{minipage}[b]{0.5\linewidth} % Adjust width as needed

        \includegraphics[width=1\linewidth]{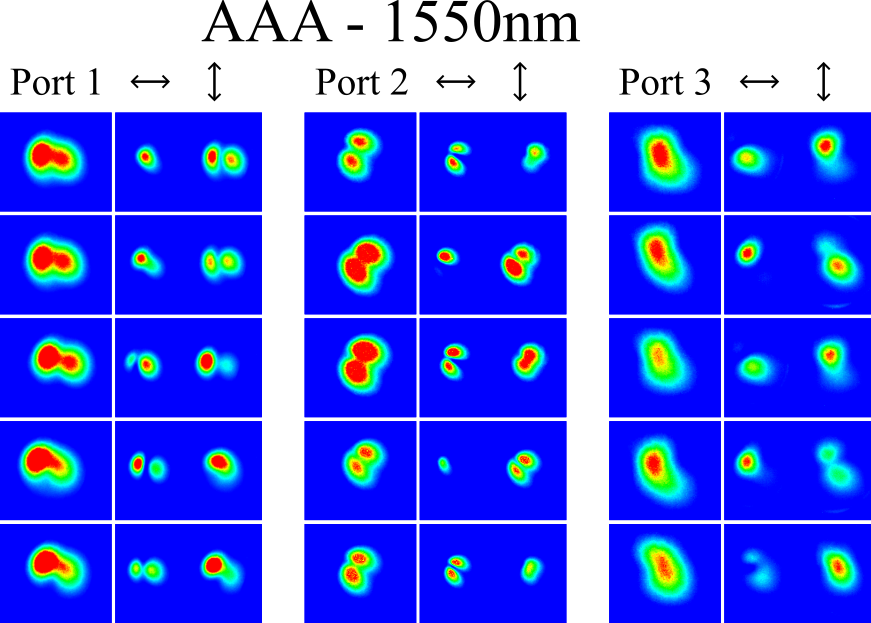}
        \caption{Optical profile response at five different polarizations of a standard \gls{pl} at 1550~\unit{nm}.}\label{fig:AAA1550nm}

        \end{minipage}
        \hfill % Adds horizontal space between minipages
        \begin{minipage}[b]{0.45\linewidth} % Adjust width as needed
\begin{center}
\setlength{\tabcolsep}{1pt}
\tiny
\begin{tabular}{ |c c|c c|c c| } 
 \hline
 \multicolumn{2}{|c|}{Port 1} & \multicolumn{2}{|c|}{Port 2} & \multicolumn{2}{|c|}{Port 3} \\

\hline\hline

PER [dB] & lin PER [dB] & PER [dB] & lin PER [dB] &PER [dB] & lin PER [dB] \\ 
$\psi$ [\degree] & DOP [\%] & $\psi$ [\degree] & DOP [\%] & $\psi$ [\degree] & DOP [\%] \\
\hline

N.A. & N.A. & N.A. & N.A. & 25.25 & 14.79 \\ 
N.A. & N.A. & N.A. & N.A. & -3.85 & 96.97 \\
\hline
N.A. & N.A. & N.A. & N.A. & 30.38 & 10.80 \\ 
N.A. & N.A. & N.A. & N.A. & 88.93 & 91.57 \\
\hline
N.A. & N.A. & N.A. & N.A. & N.A. & N.A. \\ 
N.A. & N.A. & N.A. & N.A. & N.A. & N.A. \\
\hline

N.A. & N.A. & N.A. & N.A. & N.A. & N.A. \\ 
N.A. & N.A. & N.A. & N.A. & N.A. & N.A. \\
\hline

N.A. & N.A. & N.A. & N.A. & N.A. & N.A. \\ 
N.A. & N.A. & N.A. & N.A. & N.A. & N.A. \\
\hline

\end{tabular}

        \captionof{table}{Measurements of a \gls{pl} for each port across five polarizations at 1550~\unit{nm}.}
        \label{tab:T2}
\end{center}
        \end{minipage}
    \end{figure}

\begin{figure}[h!]
        \begin{minipage}[b]{0.55\linewidth} % Adjust width as needed

        \includegraphics[width=1\linewidth]{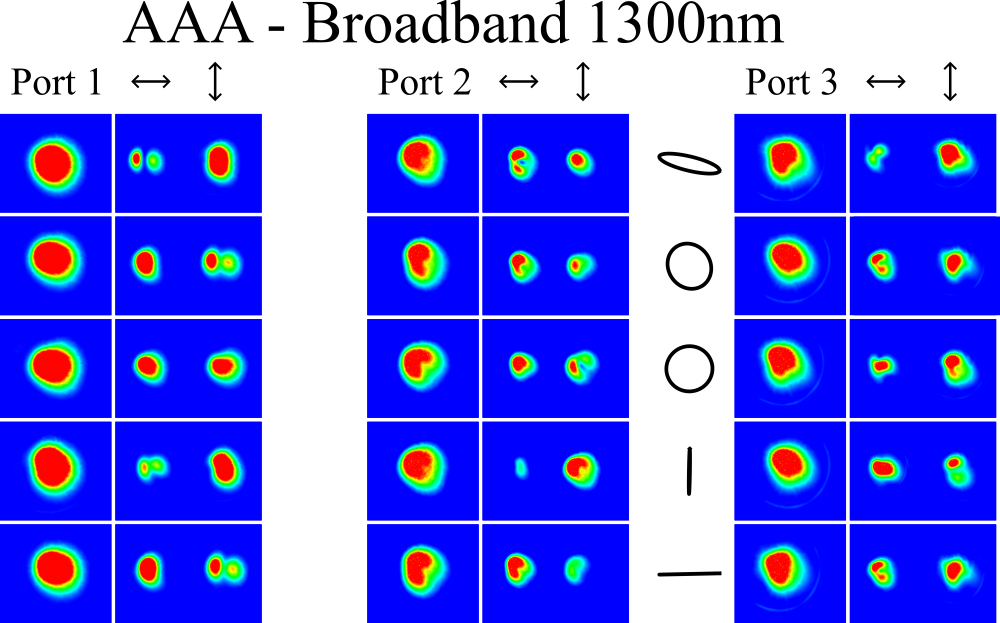}
        \caption{Optical profile response at five different polarizations of a standard \gls{pl} using a broadband source centered at 1300~\unit{nm}.}\label{fig:AAABnm}

        \end{minipage}
        \hfill % Adds horizontal space between minipages
        \begin{minipage}[b]{0.45\linewidth} % Adjust width as needed
\begin{center}
\setlength{\tabcolsep}{1pt}
\tiny
\begin{tabular}{ |c c|c c|c c| } 
 \hline
 \multicolumn{2}{|c|}{Port 1} & \multicolumn{2}{|c|}{Port 2} & \multicolumn{2}{|c|}{Port 3} \\

\hline\hline

PER [dB] & lin PER [dB] & PER [dB] & lin PER [dB] &PER [dB] & lin PER [dB] \\ 
$\psi$ [\degree] & DOP [\%] & $\psi$ [\degree] & DOP [\%] & $\psi$ [\degree] & DOP [\%] \\
\hline

N.A. & N.A.  & N.A. & N.A. & 32.48 & 3.32 \\ 
N.A. & N.A.  & N.A. & N.A. & 89.06 & 53.52 \\
\hline

N.A. & N.A.  & N.A. & N.A. & 54.99 & 4.48 \\ 
N.A. & N.A.  & N.A. & N.A. & 1.20 & 64.38 \\
\hline
N.A. & N.A.  & N.A. & N.A. & 12.79 & 6.42 \\ 
N.A. & N.A.  & N.A. & N.A. & -13.32 & 81.47 \\
\hline

N.A. & N.A.  & N.A. & N.A. & 0.24 & 0.10 \\ 
N.A. & N.A.  & N.A. & N.A. & 8.04 & 43.86 \\
\hline

N.A. & N.A.  & N.A. & N.A. & 0.95 & 0.33 \\ 
N.A. & N.A.  & N.A. & N.A. & -52.16 & 37.21 \\
\hline

\end{tabular}

        \captionof{table}{Measurements of a \gls{pl} for each port across five polarizations using a broadband source centered at 1300~\unit{nm}.}
        \label{tab:T3}
\end{center}
        \end{minipage}
    \end{figure}

Figure \ref{fig:AAA1300nm}, \ref{fig:AAA1550nm}, and \ref{fig:AAABnm} illustrate the optical response of a standard \gls{pl} with configuration AAA at wavelengths of 1300~\unit{nm}, 1550~\unit{nm}, and a broad spectrum centered at 1300~\unit{nm}, respectively. Additionally, Table \ref{tab:T1}, \ref{tab:T2}, and \ref{tab:T3} present the optical measurements for the same lantern at the same wavelengths: 1300~\unit{nm}, 1550~\unit{nm}, and broadband centered at 1300~\unit{nm}.

\newpage

\subsection{AAB configuration}

\begin{figure}[h!]
        \begin{minipage}[b]{0.55\linewidth} % Adjust width as needed

        \includegraphics[width=1\linewidth]{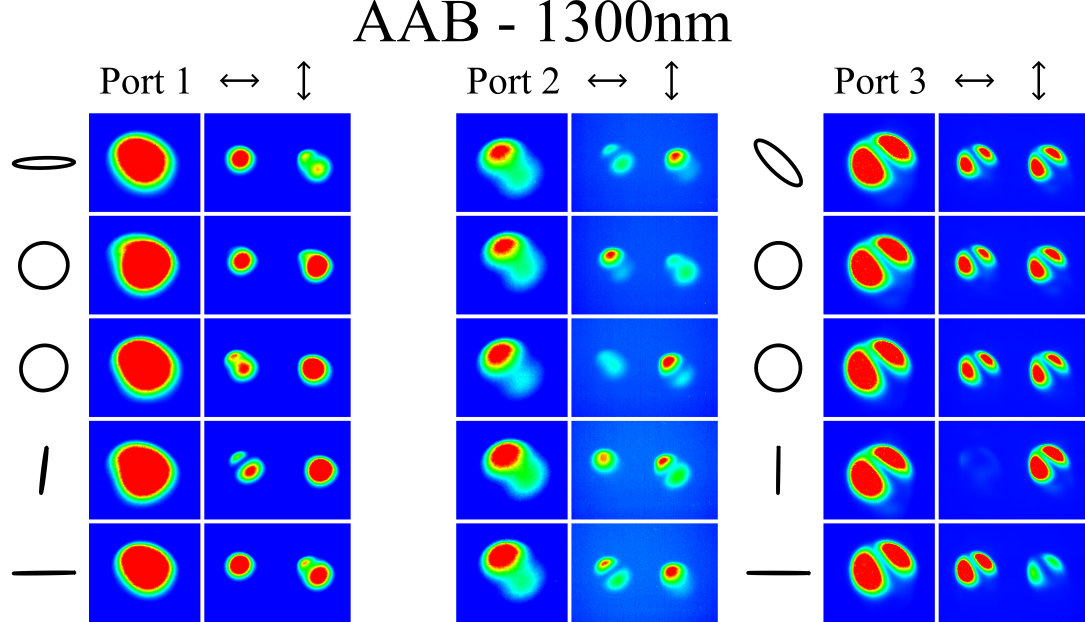}
        \caption{Optical profile response at five different polarizations of a semi-\gls{mspl} at 1300~\unit{nm}.}\label{fig:AAB1300nm}

        \end{minipage}
        \hfill % Adds horizontal space between minipages
        \begin{minipage}[b]{0.45\linewidth} % Adjust width as needed
\begin{center}
\setlength{\tabcolsep}{1pt}
\tiny
\begin{tabular}{ |c c|c c|c c| } 
 \hline
 \multicolumn{2}{|c|}{Port 1} & \multicolumn{2}{|c|}{Port 2} & \multicolumn{2}{|c|}{Port 3} \\

\hline\hline

PER [dB] & lin PER [dB] & PER [dB] & lin PER [dB] &PER [dB] & lin PER [dB] \\ 
$\psi$ [\degree] & DOP [\%] & $\psi$ [\degree] & DOP [\%] & $\psi$ [\degree] & DOP [\%] \\
\hline

15.48 & 5.53 & N.A. & N.A & 10.18 & 6.90 \\ 
1.34 & 74.12 & N.A. & N.A & -44.43 & 89.98 \\
\hline

0.59 & 0.49 & N.A. & N.A & 0.20 & 0.19 \\ 
7.79 & 83.97 & N.A. & N.A & 58.53 & 92.71 \\
\hline

0.40 & 0.32 & N.A. & N.A & 0.11 & 0.11 \\ 
33.25 & 80.80 & N.A. & N.A & 87.76 & 98.45 \\
\hline

27.57 & 3.34 & N.A. & N.A & 40.21 & 13.62 \\ 
83.00 & 54.40 & N.A. & N.A & 89.23 & 95.67 \\
\hline

35.64 & 4.91 & N.A. & N.A & 36.86 & 19.69 \\ 
0.72 & 67.71 & N.A. & N.A & -0.21 & 98.95 \\
\hline

\end{tabular}

        \captionof{table}{Measurements of a semi-\gls{mspl} for each port across five polarizations at 1300~\unit{nm}.}
        \label{tab:T4}
\end{center}
        \end{minipage}
    \end{figure}

\begin{figure}[h!]
        \begin{minipage}[b]{0.55\linewidth} % Adjust width as needed

        \includegraphics[width=1\linewidth]{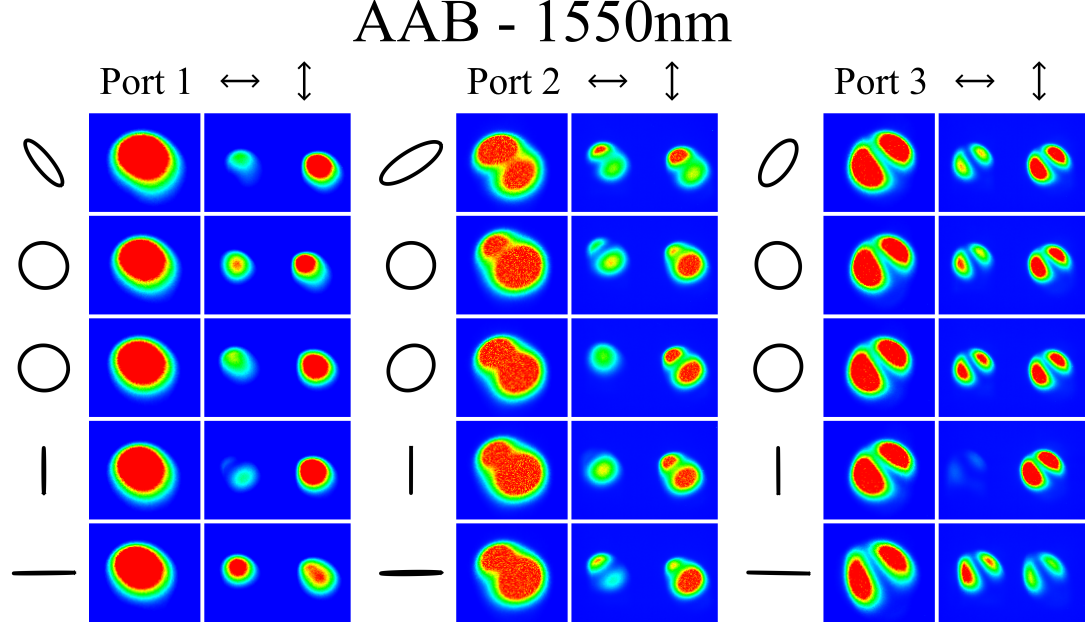}
        \caption{Optical profile response at five different polarizations of a semi-\gls{mspl} at 1550~\unit{nm}.}\label{fig:AAB1550nm}

        \end{minipage}
        \hfill % Adds horizontal space between minipages
        \begin{minipage}[b]{0.45\linewidth} % Adjust width as needed
\begin{center}
\setlength{\tabcolsep}{1pt}
\tiny
\begin{tabular}{ |c c|c c|c c| } 
 \hline
 \multicolumn{2}{|c|}{Port 1} & \multicolumn{2}{|c|}{Port 2} & \multicolumn{2}{|c|}{Port 3} \\

\hline\hline

PER [dB] & lin PER [dB] & PER [dB] & lin PER [dB] &PER [dB] & lin PER [dB] \\ 
$\psi$ [\degree] & DOP [\%] & $\psi$ [\degree] & DOP [\%] & $\psi$ [\degree] & DOP [\%] \\
\hline

11.06 & 8.73 & 9.67 & 9.67 & 6.71 & 4.67 \\ 
-51.18 & 91.83 & 32.13 & 100 & 53.68 & 83.80 \\
\hline

0.64 & 0.56 & 0.32 & 0.26 & 0.53 & 0.41 \\ 
-23.85 & 88.90 & 8.60 & 83.47 & -57.00 & 77.96 \\
\hline

0.73 & 0.71 & 1.30 & 0.93 & 0.64 & 0.55 \\ 
-7.55 & 96.84 & 42.67 & 74.75 & 31.87 & 87.10 \\
\hline

31.79 & 9.37 & 45.94 & 4.05 & 42.83 & 42.83 \\ 
-89.88 & 88.50 & 89.89 & 60.68 & -89.69 & 100 \\
\hline

35.98 & 15.77 & 27.44 & 5.87 & 41.57 & 41.57 \\ 
0.61 & 97.38 & 0.74 & 74.26 & -0.99 & 100 \\
\hline

\end{tabular}

        \captionof{table}{Measurements of a semi-\gls{mspl} for each port across five polarizations at 1550~\unit{nm}.}
        \label{tab:T5}
\end{center}
        \end{minipage}
    \end{figure}

\begin{figure}[h!]
        \begin{minipage}[b]{0.55\linewidth} % Adjust width as needed

        \includegraphics[width=1\linewidth]{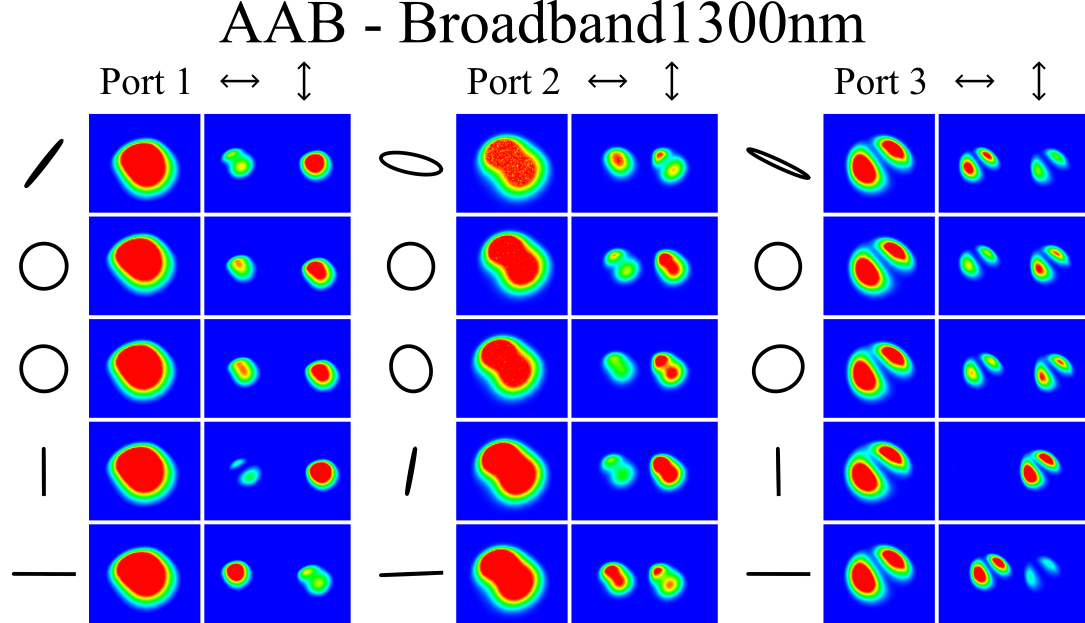}
        \caption{Optical profile response at five different polarizations of a semi-\gls{mspl} using a broadband source centered at 1300~\unit{nm}.}\label{fig:AABBnm}

        \end{minipage}
        \hfill % Adds horizontal space between minipages
        \begin{minipage}[b]{0.45\linewidth} % Adjust width as needed
\begin{center}
\setlength{\tabcolsep}{1pt}
\tiny
\begin{tabular}{ |c c|c c|c c| } 
 \hline
 \multicolumn{2}{|c|}{Port 1} & \multicolumn{2}{|c|}{Port 2} & \multicolumn{2}{|c|}{Port 3} \\

\hline\hline

PER [dB] & lin PER [dB] & PER [dB] & lin PER [dB] &PER [dB] & lin PER [dB] \\ 
$\psi$ [\degree] & DOP [\%] & $\psi$ [\degree] & DOP [\%] & $\psi$ [\degree] & DOP [\%] \\
\hline

24.40 & 5.98 & 10.41 & 3.66 & 19.52 & 13.59 \\ 
51.75 & 75.03 & -11.49 & 61.67 & -24.84 & 95.17 \\
\hline

0.16 & 0.15 & 0.21 & 0.17 & 0.25 & 0.23 \\ 
5.71 & 90.43 & -51.32 & 82.07 & -68.38 & 91.26 \\
\hline

0.16 & 0.14 & 1.42 & 0.72 & 1.19 & 1.07 \\ 
-33.74 & 93.31 & -67.54 & 54.72 & 25.01 & 90.81 \\
\hline

43.54 & 5.28 & 27.08 & 1.86 & 42.74 & 15.58 \\ 
-89.77 & 70.34 & 79.75 & 34.84 & -88.99 & 97.24 \\
\hline

58.42 & 3.94 & 44.73 & 3.70 & 53.51 & 15.16 \\ 
-0.24 & 59.59 & 2.29 & 57.34 & -0.16 & 96.95 \\
\hline

\end{tabular}

        \captionof{table}{Measurements of a semi-\gls{mspl} for each port across five polarizations using a broadband source centered at 1300~\unit{nm}.}
        \label{tab:T6}
\end{center}
        \end{minipage}
    \end{figure}

Figure \ref{fig:AAB1300nm}, \ref{fig:AAB1550nm}, and \ref{fig:AABBnm} show the optical response of a semi-\gls{mspl} configured as AAB  at wavelengths of 1300~\unit{nm}, 1550~\unit{nm}, and a broad spectrum centered around 1300~\unit{nm}, respectively. Furthermore, Table \ref{tab:T4}, \ref{tab:T5}, and \ref{tab:T6} display the measurements for the same lantern at the identical wavelengths: 1300~\unit{nm}, 1550~\unit{nm}, and broadband centered at 1300~\unit{nm}.

\newpage

\subsection{ABB configuration}

\begin{figure}[h!]
        \begin{minipage}[b]{0.55\linewidth} % Adjust width as needed

        \includegraphics[width=1\linewidth]{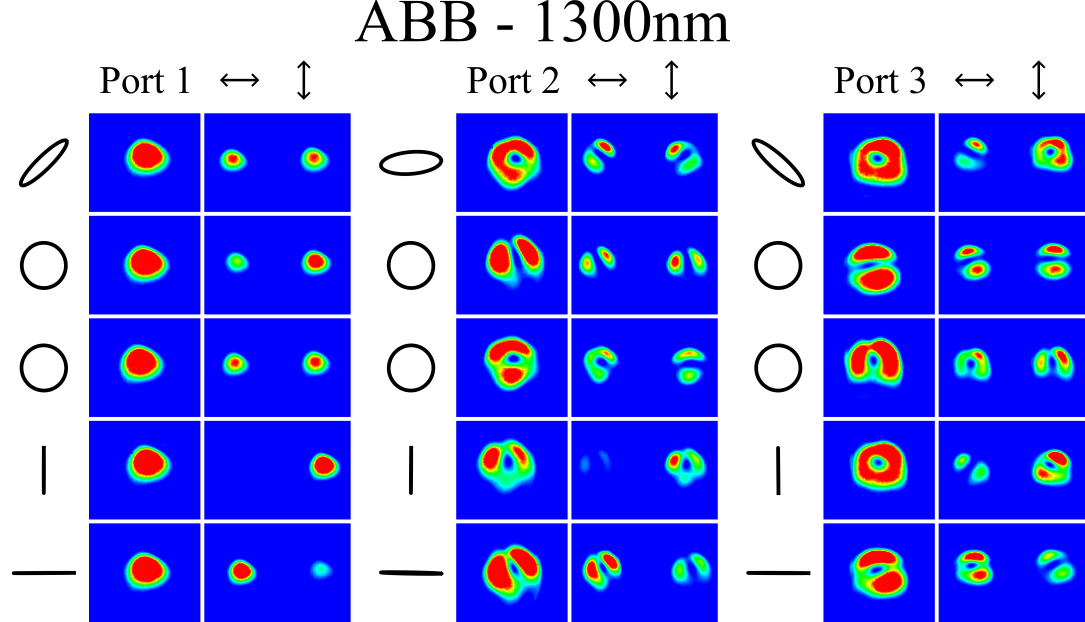}
        \caption{Optical profile response at five different polarizations of a group-\gls{mspl} at 1300~\unit{nm}.}\label{fig:ABB1300nm}

        \end{minipage}
        \hfill % Adds horizontal space between minipages
        \begin{minipage}[b]{0.45\linewidth} % Adjust width as needed
\begin{center}
\setlength{\tabcolsep}{1pt}
\tiny
\begin{tabular}{ |c c|c c|c c| } 
 \hline
 \multicolumn{2}{|c|}{Port 1} & \multicolumn{2}{|c|}{Port 2} & \multicolumn{2}{|c|}{Port 3} \\

\hline\hline

PER [dB] & lin PER [dB] & PER [dB] & lin PER [dB] &PER [dB] & lin PER [dB] \\ 
$\psi$ [\degree] & DOP [\%] & $\psi$ [\degree] & DOP [\%] & $\psi$ [\degree] & DOP [\%] \\
\hline

13.33 & 13.33 & 8.89 & 8.50 & 12.76 & 11.23 \\ 
44.16 & 100 & 4.87 & 98.61 & -41.82 & 97.64 \\
\hline

0.17 & 0.16 & 0.14 & 0.14 & 0.09 & 0.09 \\ 
-86.03 & 96.12 & -55.82 & 95.08 & -74.14 & 94.34 \\
\hline

0.07 & 0.07 & 0.08 & 0.08 & 0.13 & 0.13 \\ 
-72.01 & 99.94 & 65.69 & 94.41 & 82.84 & 100 \\
\hline

 43.54 & 14.99 & 44.91 & 7.44 & 49.75 & 11.28 \\ 
89.95 & 96.83 & 89.68 & 81.97 & -89.47 & 92.55 \\
\hline

46.05 & 46.05 & 32.80 & 4.26 & 43.06 & 19.37 \\ 
0.28 & 100 & -1.07 & 62.54 & -0.22 & 98.85 \\
\hline

\end{tabular}

        \captionof{table}{Measurements of a group-\gls{mspl} for each port across five polarizations at 1300~\unit{nm}.}
        \label{tab:T7}
\end{center}
        \end{minipage}
    \end{figure}

\begin{figure}[h!]
        \begin{minipage}[b]{0.55\linewidth} % Adjust width as needed

        \includegraphics[width=1\linewidth]{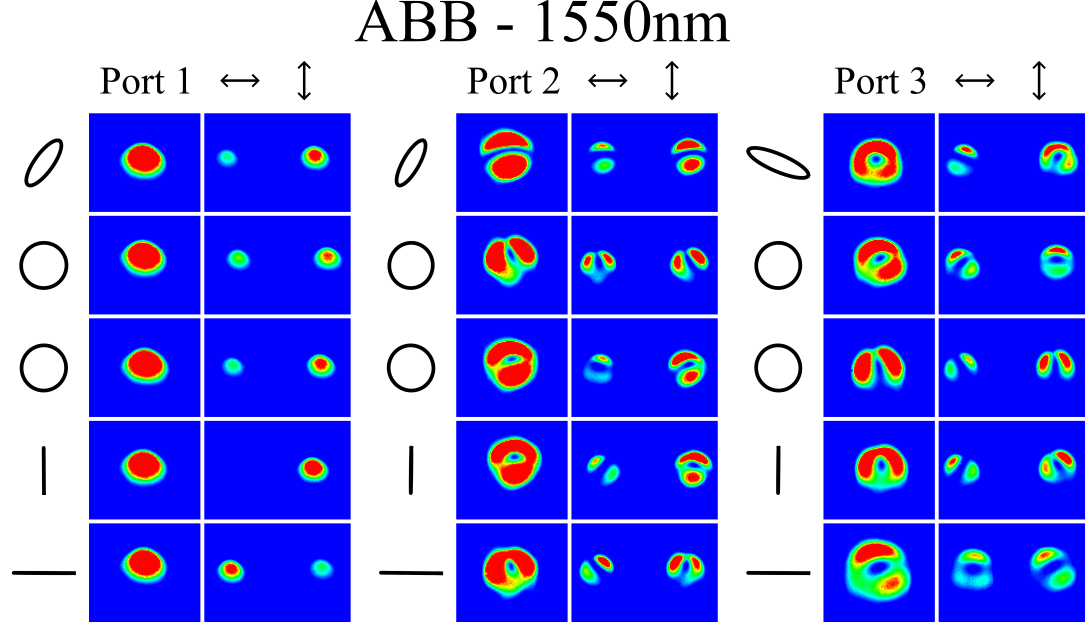}
        \caption{Optical profile response at five different polarizations of a group-\gls{mspl} at 1550~\unit{nm}.}\label{fig:ABB1550nm}

        \end{minipage}
        \hfill % Adds horizontal space between minipages
        \begin{minipage}[b]{0.45\linewidth} % Adjust width as needed
\begin{center}
\setlength{\tabcolsep}{1pt}
\tiny
\begin{tabular}{ |c c|c c|c c| } 
 \hline
 \multicolumn{2}{|c|}{Port 1} & \multicolumn{2}{|c|}{Port 2} & \multicolumn{2}{|c|}{Port 3} \\

\hline\hline

PER [dB] & lin PER [dB] & PER [dB] & lin PER [dB] &PER [dB] & lin PER [dB] \\ 
$\psi$ [\degree] & DOP [\%] & $\psi$ [\degree] & DOP [\%] & $\psi$ [\degree] & DOP [\%] \\
\hline
9.31 & 9.31 & 9.99 & 9.99 & 11.60 & 11.35 \\ 
53.94 & 100 & 60.09 & 100 & -22.90 & 99.56 \\
\hline

0.10 & 0.09 & 0.30 & 0.29 & 0.18 & 0.16 \\ 
19.60 & 96.71 & 81.39 & 95.12 & 80.84 & 89.96 \\
\hline

0.09 & 0.09 & 0.12 & 0.11 & 0.18 & 0.17 \\ 
72.73 & 99.88 & -52.02 & 92.75 & -69.92 & 97.92 \\
\hline

52.75 & 29.31 & 38.57 & 11.78 & 41.85 & 9.49 \\ 
-89.63 & 99.88 & 89.35 & 93.38 & 89.45 & 88.76 \\
\hline

48.28 & 48.28 & 74.66 & 13.05 & 51.51 & 15.76 \\ 
-0.27 & 100 & -0.57 & 94.88 & -0.26 & 97.35 \\
\hline

\end{tabular}

        \captionof{table}{Measurements of a group-\gls{mspl} for each port across five polarizations at 1550~\unit{nm}.}
        \label{tab:T8}
\end{center}
        \end{minipage}
    \end{figure}

\begin{figure}[h!]
        \begin{minipage}[b]{0.55\linewidth} % Adjust width as needed

        \includegraphics[width=1\linewidth]{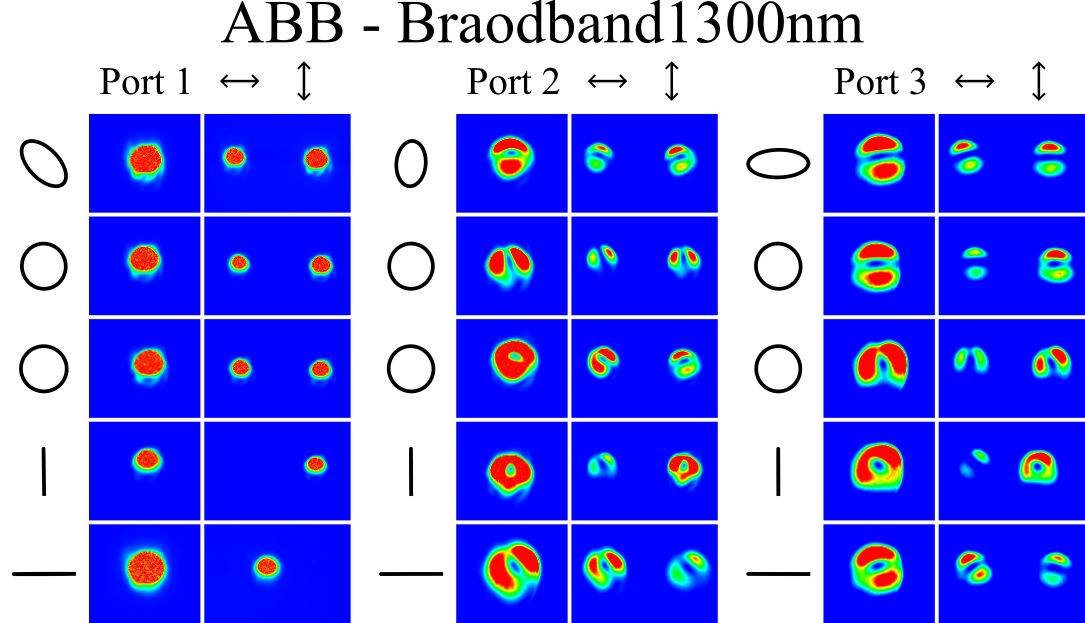}
        \caption{Optical profile response at five different polarizations of a group-\gls{mspl} using a broadband source centered at 1300~\unit{nm}.}\label{fig:ABBBnm}

        \end{minipage}
        \hfill % Adds horizontal space between minipages
        \begin{minipage}[b]{0.45\linewidth} % Adjust width as needed
\begin{center}
\setlength{\tabcolsep}{1pt}
\tiny
\begin{tabular}{ |c c|c c|c c| } 
 \hline
 \multicolumn{2}{|c|}{Port 1} & \multicolumn{2}{|c|}{Port 2} & \multicolumn{2}{|c|}{Port 3} \\

\hline\hline

PER [dB] & lin PER [dB] & PER [dB] & lin PER [dB] &PER [dB] & lin PER [dB] \\ 
$\psi$ [\degree] & DOP [\%] & $\psi$ [\degree] & DOP [\%] & $\psi$ [\degree] & DOP [\%] \\
\hline
5.34 & 4.78 & 3.73 & 3.32 & 6.86 & 6.10 \\ 
-45.60 & 94.33 & 85.10 & 92.68 & 1.08 & 95.05 \\
\hline

0.23 & 0.22 & 0.24 & 0.19 & 0.20 & 0.18 \\ 
-76.75 & 99.55 & 88.18 & 78.95 & -50.95 & 94.12 \\
\hline

0.17 & 0.17 & 0.14 & 0.11 & 0.16 & 0.16 \\ 
-10.18 & 100 & -51.05 & 80.01 & -84.69 & 100 \\
\hline

52.18 & 11.70 & 44.41 & 1.78 & 47.69 & 16.98 \\ 
-89.25 & 93.24 & -89.78 & 33.64 & -89.84 & 98.00 \\
\hline

44.12 & 9.55 & 58.46 & 19.65 & 49.70 & 20.18 \\ 
0.22 & 88.92 & 0.07 & 98.92 & -0.10 & 98.69 \\
\hline

\end{tabular}

        \captionof{table}{Measurements of a group-\gls{mspl} for each port across five polarizations using a broadband source centered at 1300~\unit{nm}.}
        \label{tab:T9}
\end{center}
        \end{minipage}
    \end{figure}

Figure \ref{fig:ABB1300nm}, \ref{fig:ABB1550nm}, and \ref{fig:ABBBnm} illustrate the optical response of a group-\gls{mspl} set up in the ABB configuration at wavelengths of 1300~\unit{nm}, 1550~\unit{nm}, and a broad spectrum centered around 1300~\unit{nm}, respectively. Additionally, Table \ref{tab:T7}, \ref{tab:T8}, and \ref{tab:T9} present the measurements for the same lantern at the corresponding wavelengths: 1300~\unit{nm}, 1550~\unit{nm}, and the broadband spectrum centered at 1300~\unit{nm}.

\newpage
\subsection{ABC configuration}

\begin{figure}[h!]
        \begin{minipage}[b]{0.55\linewidth} % Adjust width as needed

        \includegraphics[width=1\linewidth]{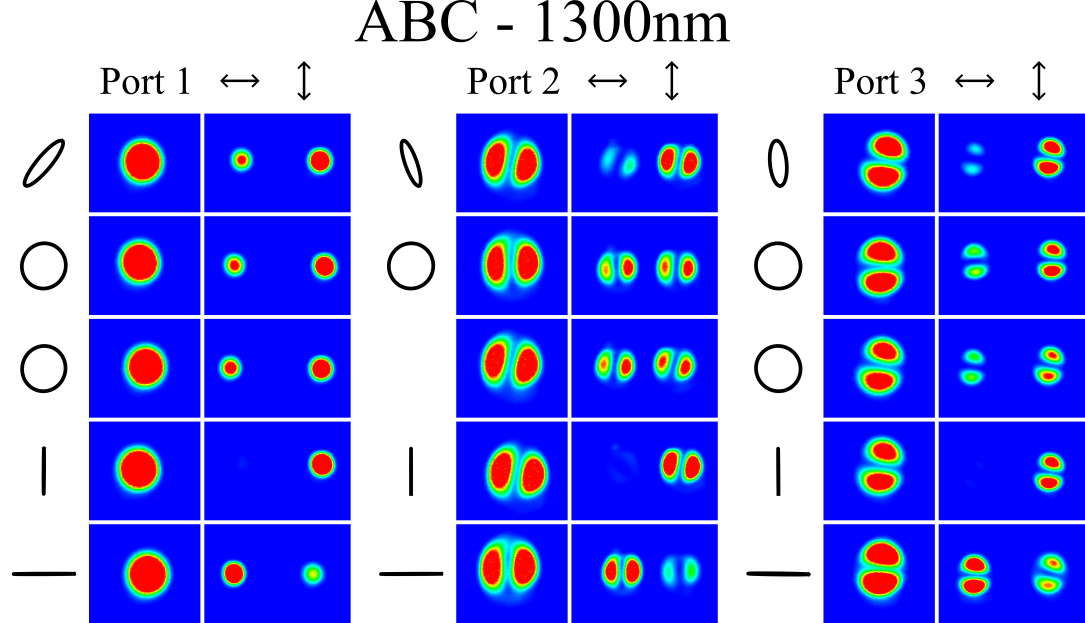}
        \caption{Optical profile response at five different polarizations of a \gls{mspl} at 1300~\unit{nm}.}\label{fig:ABC1300nm}

        \end{minipage}
        \hfill % Adds horizontal space between minipages
        \begin{minipage}[b]{0.45\linewidth} % Adjust width as needed
\begin{center}
\setlength{\tabcolsep}{1pt}
\tiny
\begin{tabular}{ |c c|c c|c c| } 
 \hline
 \multicolumn{2}{|c|}{Port 1} & \multicolumn{2}{|c|}{Port 2} & \multicolumn{2}{|c|}{Port 3} \\

\hline\hline

PER [dB] & lin PER [dB] & PER [dB] & lin PER [dB] &PER [dB] & lin PER [dB] \\ 
$\psi$ [\degree] & DOP [\%] & $\psi$ [\degree] & DOP [\%] & $\psi$ [\degree] & DOP [\%] \\
\hline

12.21 & 7.64 & 12.07 & 10.39 & 8.67 & 7.41 \\ 
49.92 & 88.10 & -70.63 & 96.88 & -84.57 & 94.72 \\
\hline

0.29 & 0.22 & 0.13 & 0.12 & 0.19 & 0.18 \\ 
64.95 & 77.30 & 67.10 & 95.69 & -54.80 & 92.85 \\
\hline

0.36 & 0.33 & N.A & N.A. & 0.17 & 0.16 \\ 
72.61 & 89.89 & N.A. & N.A. & -57.55 & 94.29 \\
\hline

 37.82 & 19.90 & 47.45 & 16.80 & 44.93 & 14.20 \\ 
89.65 & 98.99 & -89.97 & 97.91 & -89.62 & 96.20 \\
\hline

39.99 & 7.30 & 43.80 & 43.80 & 34.63 & 9.31 \\ 
0.21 & 81.41 & 0.26 & 100 & -0.68 & 88.31 \\
\hline

\end{tabular}

        \captionof{table}{Measurements of a \gls{mspl} for each port across five polarizations at 1300~\unit{nm}.}
        \label{tab:T10}
\end{center}
        \end{minipage}
    \end{figure}

\begin{figure}[h!]
        \begin{minipage}[b]{0.55\linewidth} % Adjust width as needed

        \includegraphics[width=1\linewidth]{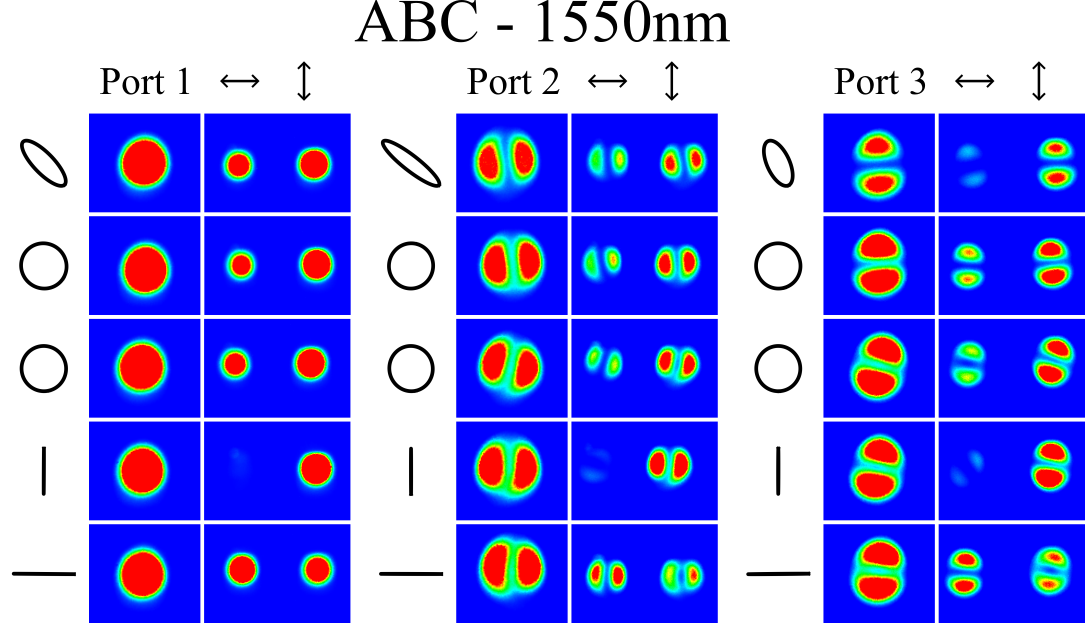}
        \caption{Optical profile response at five different polarizations of a \gls{mspl} at 1550~\unit{nm}.}\label{fig:ABC1550nm}

        \end{minipage}
        \hfill % Adds horizontal space between minipages
        \begin{minipage}[b]{0.45\linewidth} % Adjust width as needed
\begin{center}
\setlength{\tabcolsep}{1pt}
\tiny
\begin{tabular}{ |c c|c c|c c| } 
 \hline
 \multicolumn{2}{|c|}{Port 1} & \multicolumn{2}{|c|}{Port 2} & \multicolumn{2}{|c|}{Port 3} \\

\hline\hline

PER [dB] & lin PER [dB] & PER [dB] & lin PER [dB] &PER [dB] & lin PER [dB] \\ 
$\psi$ [\degree] & DOP [\%] & $\psi$ [\degree] & DOP [\%] & $\psi$ [\degree] & DOP [\%] \\
\hline

9.38 & 9.38 & 13.36 & 10.77 & 6.66 & 6.28 \\ 
-45.83 & 100 & -37.76 & 96.06 & -65.06 & 97.50 \\
\hline

0.25 & 0.23 & 0.16 & 0.15 & 0.11 & 0.11 \\ 
-31.60 & 91.91 & 73.71 & 96.61 & 73.47 & 96.41 \\
\hline

0.08 & 0.08 & 0.31 & 0.31 & 0.16 & 0.15 \\ 
-76.02 & 96.70 & -77.82 & 99.56 & 87.39 & 98.04 \\
\hline

45.42 & 12.35 & 43.37 & 43.37 & 40.55 & 40.55 \\ 
89.72 & 94.18 & -89.96 & 100 & 89.90 & 100 \\
\hline

50.99 & 26.53 & 48.50 & 48.50 & 44.74 & 44.74 \\ 
-0.43 & 99.78 & -0.15 & 100 & 0.48 & 100 \\
\hline

\end{tabular}

        \captionof{table}{Measurements of a \gls{mspl} for each port across five polarizations at 1550~\unit{nm}.}
        \label{tab:T11}
\end{center}
        \end{minipage}
    \end{figure}

\begin{figure}[h!]
        \begin{minipage}[b]{0.55\linewidth} % Adjust width as needed

        \includegraphics[width=1\linewidth]{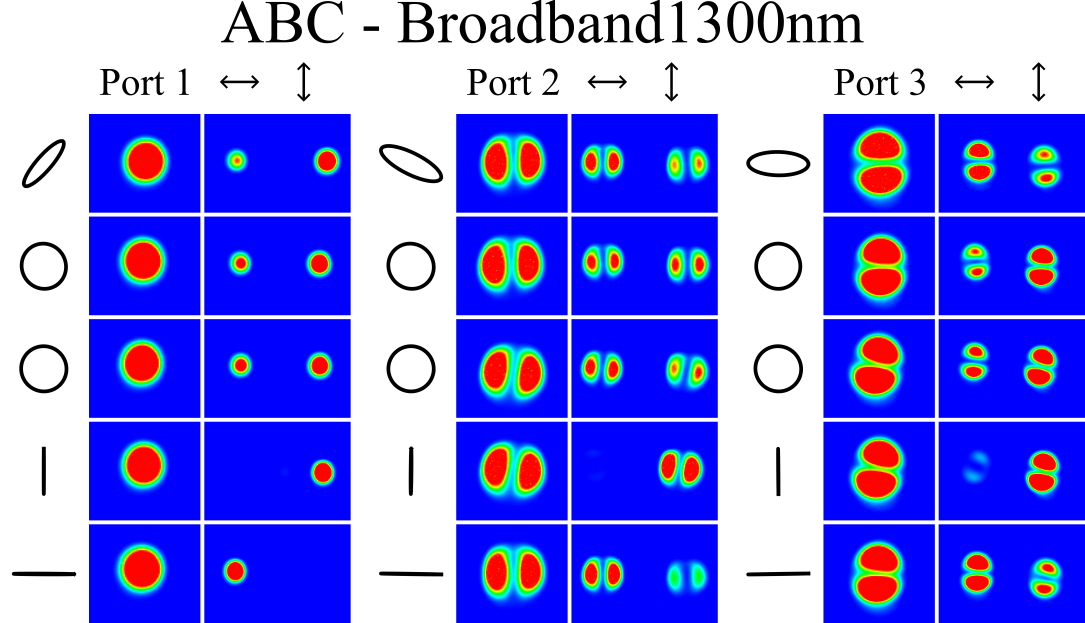}
        \caption{Optical profile response at five different polarizations of a \gls{mspl} using a broadband source centered at 1300~\unit{nm}.}\label{fig:ABCBnm}

        \end{minipage}
        \hfill % Adds horizontal space between minipages
        \begin{minipage}[b]{0.45\linewidth} % Adjust width as needed
\begin{center}
\setlength{\tabcolsep}{1pt}
\tiny
\begin{tabular}{ |c c|c c|c c| } 
 \hline
 \multicolumn{2}{|c|}{Port 1} & \multicolumn{2}{|c|}{Port 2} & \multicolumn{2}{|c|}{Port 3} \\

\hline\hline

PER [dB] & lin PER [dB] & PER [dB] & lin PER [dB] &PER [dB] & lin PER [dB] \\ 
$\psi$ [\degree] & DOP [\%] & $\psi$ [\degree] & DOP [\%] & $\psi$ [\degree] & DOP [\%] \\
\hline
11.86 & 5.02 & 9.91 & 5.24 & 8.29 & 5.63 \\ 
48.79 & 74.52 & -26.39 & 78.02 & -0.97 & 85.27 \\
\hline

0.28 & 0.19 & 0.37 & 0.30 & 0.15 & 0.12 \\ 
-55.39 & 70.76 & -37.11 & 82.63 & -83.16 & 81.80 \\
\hline

0.12 & 0.08 & 0.09 & 0.08 & 0.22 & 0.17 \\ 
-21.96 & 72.64 & -6.20 & 84.96 & -21.39 & 77.77 \\
\hline

41.65 & 33.00 & 34.73 & 1.02 & 43.71 & 11.48 \\ 
89.95 & 79.89 & 89.50 & 20.99 & -89.55 & 92.90 \\
\hline

37.79 & 7.61 & 42.37 & 42.37 & 46.87 & 7.36 \\ 
-0.29 & 82.10 & -0.84 & 100 & 0.90 & 81.64 \\
\hline

\end{tabular}

        \captionof{table}{Measurements of a \gls{mspl} for each port across five polarizations using a broadband source centered at 1300~\unit{nm}.}
        \label{tab:T12}
\end{center}
        \end{minipage}
    \end{figure}

Figures \ref{fig:ABC1300nm}, \ref{fig:ABC1550nm}, and \ref{fig:ABCBnm} display the optical response of a \gls{mspl} with configuration ABC at wavelengths of 1300~\unit{nm}, 1550~\unit{nm}, and across a broad spectrum centered at 1300~\unit{nm}, respectively. Moreover, Tables \ref{tab:T10}, \ref{tab:T11}, and \ref{tab:T12} provide the optical measurements for the same lantern at those wavelengths: 1300~\unit{nm}, 1550~\unit{nm}, and broadband centered at 1300~\unit{nm}.

\bibliographystyle{unsrt}

\bibliography{sample}

\end{document}